\documentclass[twocolumn,noshowpacs,amsmath,amssymb]{revtex4-2}
\usepackage{graphicx}
\usepackage{amsfonts}
\usepackage{dcolumn}
\usepackage{color}
\usepackage{CJK}
\usepackage{hyperref}

\begin{document}
	
\newcommand{\gin}[1]{{\bf\color{blue}#1}}
\def\bc{\begin{center}}
\def\ec{\end{center}}
\def\bea{\begin{eqnarray}}
\def\eea{\end{eqnarray}}
\newcommand{\avg}[1]{\langle{#1}\rangle}
\newcommand{\Avg}[1]{\left\langle{#1}\right\rangle}
\newcommand{\bme}[1]{\boldsymbol{#1}}	
	
\title{First passage of a diffusing particle under stochastic resetting in bounded domains with spherical symmetry}
\author{Hanshuang Chen$^{1}$}\email{chenhshf@ahu.edu.cn}
\author{Feng Huang{$^{2,3}$}}
	
\affiliation{$^{1}$School of Physics and Optoelectronics Engineering, Anhui
	University, Hefei 230601, China \\ $^2$Key Laboratory of Advanced Electronic Materials and Devices \& School of Mathematics and Physics, Anhui Jianzhu University, Hefei, 230601, China \\ $^3$Key Laboratory of Architectural Acoustic Environment of Anhui Higher Education Institutes, Hefei, 230601, China}

\date{\today}
	
\begin{abstract}	
We investigate the first passage properties of a Brownian particle diffusing freely inside a $d$-dimensional sphere with absorbing spherical surface subject to stochastic resetting. We derive the mean time to absorption (MTA) as functions of resetting rate $\gamma$ and initial distance $r$ of the particle to the centre of the sphere. We find that when $r>r_c$ there exists a nonzero optimal resetting rate $\gamma_{{\rm opt}}$ at which the MTA is a minimum, where $r_c=\sqrt {d/\left( {d + 4} \right)} R$ and $R$ is the radius of the sphere. As $r$ increases, $\gamma_{{\rm opt}}$ exhibits a continuous transition from zero to nonzero at $r=r_c$.  Furthermore, we consider that the particle lies between two two-dimensional or three-dimensional concentric spheres with absorbing boundaries, and obtain the domain in which resetting expedites the MTA, which is $(R_1, r_{c_1}) \cup (r_{c_2},R_2)$, with $R_1$ and $R_2$ being the radii of inner and outer spheres, respectively. Interestingly, when $R_1/R_2$ is less than a critical value, $\gamma_{{\rm opt}}$ exhibits a discontinuous transition at $r=r_{c_1}$; otherwise, such a transition is continuous. However, at $r=r_{c_2}$, the transition is always continuous.

\end{abstract}
	\maketitle
\section{Introduction}	
First passage underlies a wide variety of stochastic phenomena that have broad applications to physics, chemistry, biology, and social sciences \cite{redner2001guide,van1992stochastic,klafter2011first,bray2013persistence}. Indeed, chemical and biochemical reactions \cite{RevModPhys.85.135}, foraging strategies of animals \cite{RevModPhys.83.81}, and the spread of diseases on social networks or of viruses through the world wide web \cite{RevModPhys.87.925} are often controlled by first encounter events.

Recently, first passage properties under resetting have been an active topic in the field of statistical physics (see \cite{evans2020stochastic} for a recent review). Resetting refers to a sudden interruption of a stochastic process followed by its starting anew, which finds its applications in search problems  \cite{PhysRevLett.113.220602,PhysRevE.92.052127}, the optimization of randomized computer
algorithms \cite{PhysRevLett.88.178701}, and in the field of biophysics \cite{reuveni2014role,rotbart2015michaelis}.  Surprisingly, for a simple diffusive Brownian particle, resetting renders an infinite mean first passage time finite, which can be also minimized at a specific resetting rate \cite{evans2011diffusion}. These nontrivial findings have initiated an enormous recent activities in this topic, including theory \cite{evans2011diffusion2,evans2014diffusion,pal2015diffusion,PhysRevLett.112.240601,meylahn2015large,PhysRevE.93.060102,pal2017first,chechkin2018random,PhysRevLett.120.080601,magoni2020ising,de2020optimization,PhysRevLett.122.020602,PhysRevE.101.062147,evans2018run,PhysRevE.102.052129,huang2021random}, experiments \cite{tal2020experimental,besga2020optimal}, and applications \cite{gupta2014fluctuating,fuchs2016stochastic,pal2017integral,gupta2020work,magoni2020ising} . 

Resetting can either hinder or facilitate in the completion of a stochastic process. 
There have been realizations that so-called ``resetting transition'' occurs at some parameter of the underlying model, which distinguishes the role of resetting in the first passage properties. Resetting transition can be first \cite{PhysRevLett.113.220602,PhysRevE.92.062115} or second order \cite{Christou2015,PhysRevE.97.062106} like in the classical phase transition. A Landau-like theory was also used to characterize phase transitions in resetting systems \cite{PhysRevResearch.1.032001}. An interesting question arises: what is condition under which resetting will actually expedite completion of a stochastic process? Reuveni \cite{PhysRevLett.116.170601} first made a universal observation that the relative standard deviation associated with the first passage time of an optimally restarted process is always unity. Pal and Reuveni \cite{pal2017first} further showed restart has the ability to expedite the completion of the underlying stochastic process if the following ``restart criterion'' is met,   
\begin{eqnarray}\label{eq1.5}
\rm CV=\frac{{\sqrt {\langle {\tau_0^2} \rangle  - {\langle {\tau_0} \rangle }^2} }} {\langle {\tau_0} \rangle} > 1
\end{eqnarray}  
where $\langle {\tau_0} \rangle$ and $\langle {\tau_0^2} \rangle$ are the first and second moments of the first passage time of a stochastic process without resetting, respectively. By a rearrangement of Eq.\ref{eq1.5}, the condition can be interpreted as the mean time to completion $\langle {\tau_0} \rangle$ being less than the mean residual life time $\langle {\tau^2_0} \rangle/2 \langle {\tau_0} \rangle$ \cite{Gallager2013}. Interestingly, the condition in Eq.\ref{eq1.5} was also understood by so-called ``inspection paradox'' \cite{Pal2021}. The usefulness of Eq.\ref{eq1.5} was demonstrated in systems of a Brownian walker in a one-dimensional domain without force field \cite{Durang2019,PhysRevE.99.032123,PhysRevE.103.052129} or in the presence of linear, harmonic, power-law, and logarithm potentials \cite{Ray2019,PhysRevE.99.022130,PhysRevE.103.052129,PhysRevResearch.1.032001,JCP2020.152.234110,JCP2020.153.234904,JCP2021.154.171103}, or in a high-dimensional system with a power-law potential \cite{PhysRevE.102.032145}.

However, we should stress that $\rm CV < 1$ does not necessarily imply that resetting cannot expedite the completion of a stochastic process. While in this latter case the introduction of a small resetting rate will surely increase the mean completion time, resetting with an intermediate rate may still expedite completion \cite{rotbart2015michaelis,PhysRevResearch.1.032001}.

In this work, we aim to investigate the effect of stochastic resetting on first passage properties of a freely diffusive particle confined in  spherically symmetric systems with absorbing boundary, and obtain a general condition under which resetting will expedite the completion of the diffusion process. This setting is relevant to stochastic dynamics of mesoscopic particles or macromolecules inside a confined space.   
One is the so-called narrow-escape problem \cite{Schuss_J.Sci.Comput.2012} that gives the mean time when a Brownian particle or a polymer trapped in a confined domain escapes from a single narrow opening for the first time \cite{PhysRevLett.100.168105,JCP2021.155.194102,PhysRevLett.86.3188,JCP2003.118.5174,JCP2016.145.084906,JCP2016.144.144901}. Examples include an ion finding an open ion channel situated within the cell membrane or a protein receptor locating a particular target binding site \cite{RevModPhys.85.135}. In soft condensed matter and in a variety of biological systems, the transport of Brownian particles in restricted channels is also relevant in this context \cite{JPC1992.96.3926,PhysRevE.64.061106,PhysRevE.84.011109,PhysRevE.84.011149,PhysRevE.85.031128,PhysRevE.75.051111,PhysRevE.75.061126,PhysRevE.82.032103,PhysRevLett.96.130603,PhysRevLett.101.130602,PhysRevLett.104.020601,JCP2010.133.204102,JCP2010.132.224102,JCP2010.136.114104}.

To be specific, we consider the particle  diffusing freely inside a sphere or between two concentric spheres subject to resetting at random times, but with a constant rate $\gamma$. Using a renewal approach, we derive the mean time to adsorption (MTA) as functions of $\gamma$ and the initial distance $r$ of the particle to the centre of sphere. For a $d$-dimensional sphere of radius $R$, the MTA can be optimized at a nonzero resetting rate $\gamma_{{\rm opt}}$ when $r>r_c=\sqrt{d/d+4})R$, and $\gamma_{{\rm opt}}$ shows a continuous transition at $r=r_c$. For two concentric spheres, the resetting can lead to more rich feature of phase transitions. The  domains in which the resetting expedites the MTA are $(R_1, r_{c_1}) \cup (r_{c_2},R_2)$, where $R_1$ and $R_2$ are the radii of the inner sphere and outer sphere, respectively. Interestingly, when $R_1/R_2$ is less than a critical value, $\gamma_{{\rm opt}}$ shows a discontinuous transition at $r=r_{c_1}$ and a continuous transition at $r=r_{c_2}$. Otherwise, both the transitions are continuous. Finally, the asymptotical behaviors in the limit of $R_1/R_2 \to 0$ at two transitions are shown. 
  
\section{Mean time to adsorption of a freely diffusive particle in a bounded domain}
Let us begin with a general theory for a freely diffusive particle in a bounded domain $\Omega$ with absorbing boundaries (denoted by $\partial \Omega$) in the absence of resetting. Letting $p(\bme{x}, t|\bme{x}_0)$ denote the conditional probability density of finding the particle at a position $\bme{x}$ at time $t$, provided that its initial position was $\bme{x}_0 \in \Omega$, we write down the Fokker–Planck equation for the process,
\begin{eqnarray}\label{eq1.70}
\frac{{\partial p\left( {\bme{x},t| {{\bme{x}_0}} } \right)}}{{\partial t}} = D{\nabla ^2}p\left( {\bme{x},t| {{\bme{x}_0}} } \right), \quad \bme{x} \in \Omega 
\end{eqnarray}
with boundary condition
\begin{eqnarray}\label{eq1.71}
p\left( {\bme{x},t| {{\bme{x}_0}} } \right)=0, \quad \bme{x} \in \partial\Omega, 
\end{eqnarray}
where $D$ is the diffusion coefficient. The process ends when the particle hits the absorbing boundaries. Denote by $Q_0(t|\bme{x}_0)=\int_{\bme{x} \in \Omega } {p\left( {\bme{x} ,t| {{\bme{x}_0 }} } \right)d\bme{x}} $ the survival probability that the particle has not yet been absorbed up to time $t$ in the absence of resetting providing that it has started from position $\bme{x}_0$, which satisfies a backward Fokker-Planck equation \cite{evans2020stochastic,evans2011diffusion}, 
\begin{eqnarray}\label{eq1.7}
\frac{{\partial {Q_0}(t|\bme{x}_0)}}{{\partial t}} = D{\nabla ^2}{Q_0}(t|\bme{x}_0),
\end{eqnarray}
with boundary condition
\begin{eqnarray}\label{eq1.72}
{ {Q_0}(t|\bme{x}_0)} = 0, \quad \bme{x}_0 \in \partial\Omega.
\end{eqnarray}

Performing the Laplace transform for ${ {Q_0}(t|\bme{x}_0)}$, ${{\tilde Q}_0}\left( {s|\bme{x}_0} \right) = \int_0^\infty  {{e^{ - st}}{Q_0}\left( {t|\bme{x}_0} \right)dt}$, Eq.\ref{eq1.7} becomes \cite{redner2001guide,evans2020stochastic}
\begin{eqnarray}\label{eq1.8}
s{{\tilde Q}_0}(s|\bme{x}_0) - 1 = D{\nabla ^2}{{\tilde Q}_0}(s|\bme{x}_0),
\end{eqnarray}
and the boundary condition in Eq.(\ref{eq1.72}) translates to $\tilde{Q}_0(s|\bme{x}_0)=0$ for $\bme{x}_0 \in \partial\Omega$. In Eq.(\ref{eq1.8}), we have assumed that the particle does not start from absorbing boundaries, such that the initial condition is $Q_0(0| \bme{x}_0)=1$. 

The stochastic process is terminated once the particle hits the absorbing boundaries. Letting $\tau_0$ denote the time to absorption and the probability density of $\tau_0$ is given by $-\partial { {Q_0}(t|\bme{x}_0)} / \partial t $ \cite{redner2001guide,van1992stochastic}. This allows us to calculate any moment of $\tau_0$ from $\tilde{Q}_0(s|\bme{x}_0)$ following the relation \cite{redner2001guide,JCP2020.153.234904},
\begin{eqnarray}\label{eq2.01}
\langle {{\tau^n _0}( \bme{x_0} )} \rangle  =(-1)^{n-1}  n \mathop {\lim }\limits_{s \to 0}  \frac{{{d^{n - 1}}{{\tilde Q}_0}\left( {s|\bme{x}_0} \right)}}{{d{s^{n - 1}}}}.
\end{eqnarray}
In particular, $n=1$ corresponds to the mean time to adsorption (MTA) and $n=2$ to 
the mean squared time to adsorption (MSTA).

We now explore the effect of stochastic resetting on the diffusion process. We assume that at each time the particle is reset instantaneously to a given position $\bme{x}_r$ with a constant rate $\gamma$.
Furthermore, the survival probability $Q(t|\bme{x}_0)$ in the presence of resetting  can be connected with $Q_0(t|\bme{x}_0)$ by a last renewal equation \cite{pal2016diffusion,chechkin2018random}, 
\begin{eqnarray}\label{eq1.1}
Q(t|\bme{x}_0) &=& {e^{ - \gamma t}}{Q_0}(t|\bme{x}_0) \nonumber \\ &+& \gamma \int_0^t {{e^{ - \gamma \tau }}{Q_0}( {\tau| \bme{x}_r}  )} Q( {t - \tau|\bme{x}_0 } )d\tau. 
\end{eqnarray}	
The first term in Eq.\ref{eq1.1} represents trajectories in which there has been no resetting. The second term represents trajectories in which resetting has occurred at least once. The integral is over $\tau$, the
time elapsed since the last reset and we have a convolution of survival probabilities: survival starting from $\bme{x}_0$ with resetting up to time $t-\tau$ (the time of the last reset) and survival starting
from $\bme{x}_r$ in the absence of resetting for duration $\tau$.

In the Laplace domain, Eq.\ref{eq1.1} becomes
\begin{eqnarray}\label{eq1.2}
\tilde Q({s|\bme{x}_0}) = \frac{{{{\tilde Q}_0}({\gamma+s|\bme{x}_0})}}{{1-\gamma {{\tilde Q}_0}({\gamma+s|{\bme{x}_r}} )}}.
\end{eqnarray}	
If the resetting position coincides with initial position, i.e. $\bme{x}_r = \bme{x}_0$, Eq.\ref{eq1.2} simplifies to
\begin{eqnarray}\label{eq1.3}
\tilde Q({s|\bme{x}_0}) = \frac{{{{\tilde Q}_0}({\gamma+s|\bme{x}_0})}}{{1-\gamma {{\tilde Q}_0}({\gamma+s|{\bme{x}_0}} )}}.
\end{eqnarray}	

The MTA in the presence of resetting is given by
\begin{eqnarray}\label{eq1.4}
\langle {\tau ( \bme{x}_0)} \rangle  = \tilde Q( {0|\bme{x}_0} ) = \frac{{{{\tilde Q}_0}( {\gamma | \bme{x}_0} )}}{{1-\gamma {{\tilde Q}_0}( {\gamma  | \bme{x}_0} )}},
\end{eqnarray}
where we have utilized Eq.\ref{eq1.3} in the last step. In terms of Eq.\ref{eq2.01} and Eq.\ref{eq1.4}, it is not hard to verify that the condition given in Eq.\ref{eq1.5} is equivalent to the derivative of $\langle {\tau} \rangle$ with respect to $\gamma$ being less than zero at $\gamma=0$.

\begin{figure}
	\centerline{\includegraphics*[width=0.7\columnwidth]{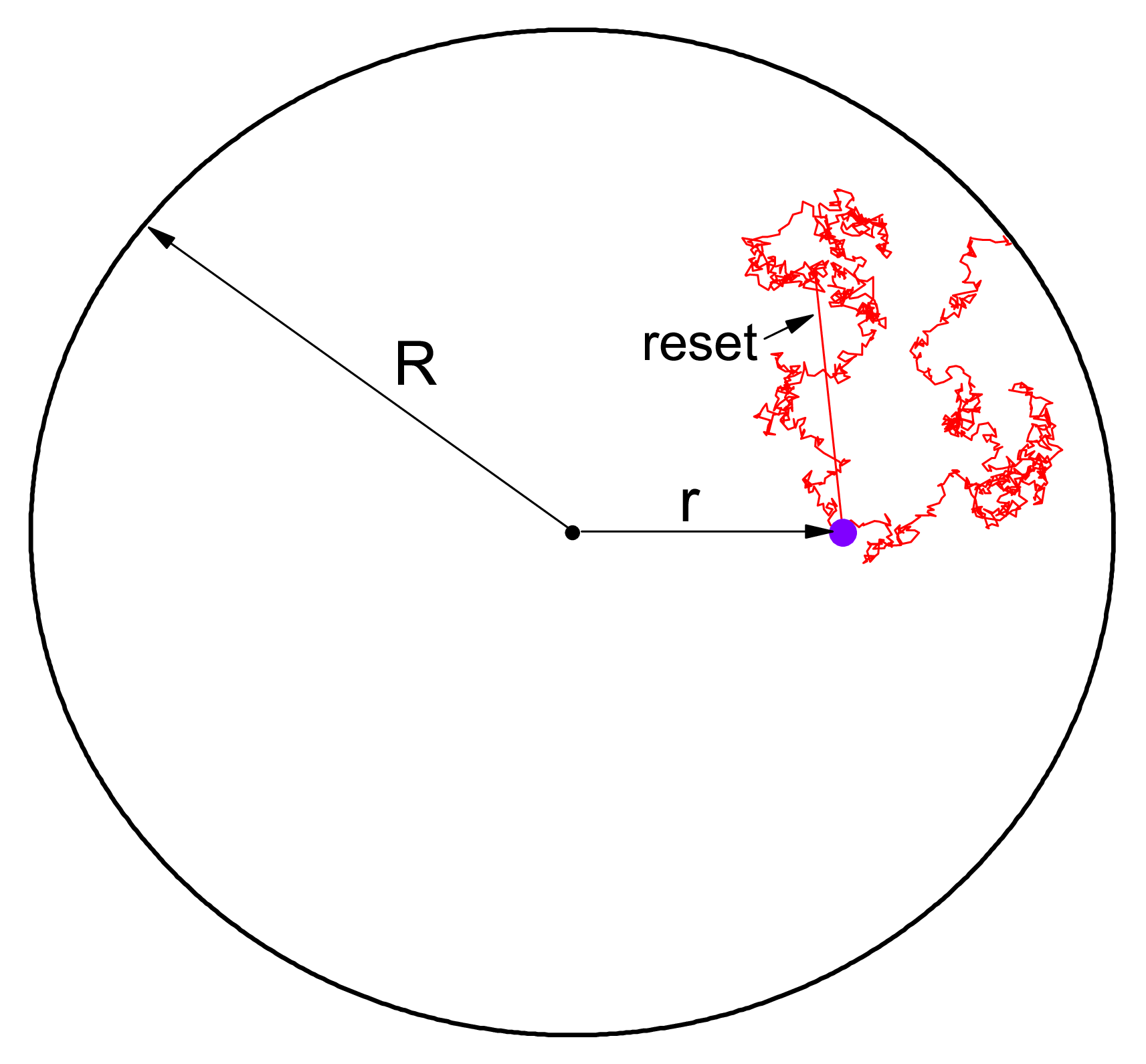}}
	\caption{A Brownian particle diffuses freely inside a $d$-dimensional sphere of radius $R$ under stochastic resetting. The particle starts from a distance $r$ to the centre of sphere. The process is terminated once the particle hits the absorbing spherical surface. } \label{fig1}
\end{figure}

\section{Results in $d$-dimensional sphere}

We consider a $d$-dimensional sphere of radius $R$, where spherical surface is an absorbing boundary and the diffusing particle starts from a distance $r$($<R$) to the centre of sphere. With the rate $\gamma$, the particle is reset to initial position. See Fig.\ref{fig1} for an illustration. Let us first derive the survival probability in the absence of resetting. Since the system has the spherical symmetry and only the radial part is relevant, Eq.\ref{eq1.8} can be written as
\begin{eqnarray}\label{eq2.1}
\frac{D}{{{r^{d - 1}}}}\frac{\partial }{{\partial r}}\left( {{r^{d - 1}}\frac{{\partial {{\tilde Q}_0}( {s|r} )}}{{\partial r}}} \right) - s{{\tilde Q}_0}( {s|r} ) + 1 = 0.
\end{eqnarray}
The general solution to Eq.\ref{eq2.1} is
\begin{eqnarray}\label{eq2.2}
{{\tilde Q}_0}( {s|r} ) &=& \frac{1}{s} +{C_1}{r^{1 - d/2}}{I_{d/2 - 1}}\left( {\alpha r} \right) \nonumber \\& +&  {C_2}{r^{1 - d/2}}{K_{d/2 - 1}}( {\alpha r} ) ,
\end{eqnarray}
where $I_m(z)$ and $K_m(z)$ are the modified Bessel functions of the first kind and of the second kind, respectively. $\alpha=\sqrt {s/D}$, $C_1$ and $C_2$ can be determined by the boundary conditions. 

Considering the following boundary conditions,
\begin{eqnarray}\label{eq2.21}
{\tilde Q}_0(s|0)< \infty, \quad {\tilde Q}_0(  s |R ) = 0.  
\end{eqnarray}
The first condition in Eq.(\ref{eq2.21}) implies that when the partical starts from the centre of sphere the survival probability is a finite value, which leads to $C_2=0$ as ${K_{d/2 - 1}}\left( {\alpha r} \right)$ diverges at $r=0$. The second condition in Eq.(\ref{eq2.21}) indicates that the survival probability equals to zero when the particle starts from absorbing spherical surface, from which we can fix the coefficient $C_1$. Therefore, ${\tilde Q}_0$ can be written as
\begin{eqnarray}\label{eq2.3}
{{\tilde Q}_0}( {s|r} ) = \frac{1}{s} - \frac{{{r^{1 - d/2}}{I_{d/2 - 1}}( {\alpha r} )}}{{s{R^{1 - d/2}}{I_{d/2 - 1}}( {\alpha R} )}}.
\end{eqnarray}

Let us define a dimensionless length (rescaled with the radius of sphere $R$) and time (rescaled with the diffusion time $R^2/D$) by setting 
\begin{eqnarray}
\bar r=r/R, \qquad  \bar \tau=\tau D/R^2.
\end{eqnarray}

The dimensionless MTA and MSTA without resetting are obtained from Eq.(\ref{eq2.01}) and Eq.(\ref{eq2.3}), given by
\begin{eqnarray}\label{eq2.42}
\langle {{\bar \tau _0}( \bar{r} )} \rangle   = \frac{{{1} - {\bar r^2}}}{{2d}},
\end{eqnarray}
and 
\begin{eqnarray}\label{eq2.43}
\langle {{\bar \tau^2 _0}( \bar{r} )} \rangle  =\frac{{\left( {1 - {{\bar r}^2}} \right)\left[ {( {d + 4} ) - d{{\bar r}^2}} \right]}}{{4{d^2}( {d + 2} )}}.
\end{eqnarray}
Substituting Eq.\ref{eq2.42} and Eq.\ref{eq2.43} into Eq.\ref{eq1.5}, we arrive at the domain in which restart expedites the MTA, 
\begin{eqnarray}\label{eq2.6}
\bar r \in \left( {\bar r_c,1} \right), \quad \bar r_c= \sqrt {\frac{d}{{d + 4}}} .
\end{eqnarray}
{Eq.(\ref{eq2.6}) is one of main theoretical results of the present work. It is shown that such a critcal distance $\bar{r}_c$ is dimension-dependent. In particular,  $d=1$ corresponds to a one-dimensional interval $(-R,R)$ with absorbing end points, in which the acceleration domain in Eq.\ref{eq2.6} becomes $(-R,-\frac{1}{\sqrt{5}}R) \cup (\frac{1}{\sqrt{5}}R,R)$, in agreement with the results of Ref.\cite{PhysRevE.97.062106,PhysRevE.99.032123,Durang2019}.

Substituting Eq.\ref{eq2.3} into Eq.\ref{eq1.4}, we obtain a dimensionless MTA in the presence of resetting,
\begin{eqnarray}\label{eq2.71}
\langle {\bar \tau } \rangle  =\frac{{{{\bar r}^{d/2 - 1}}{I_{d/2 - 1}}\left( {\sqrt {\bar \gamma } } \right)}}{{\bar \gamma {I_{d/2 - 1}}\left( {\bar r\sqrt {\bar \gamma } } \right)}} - \frac{1}{{\bar \gamma }},
\end{eqnarray}
where $\bar \gamma  = \gamma {R^2}/D$ is a dimensionless resetting rate. 

\begin{figure}
	\centerline{\includegraphics*[width=1.0\columnwidth]{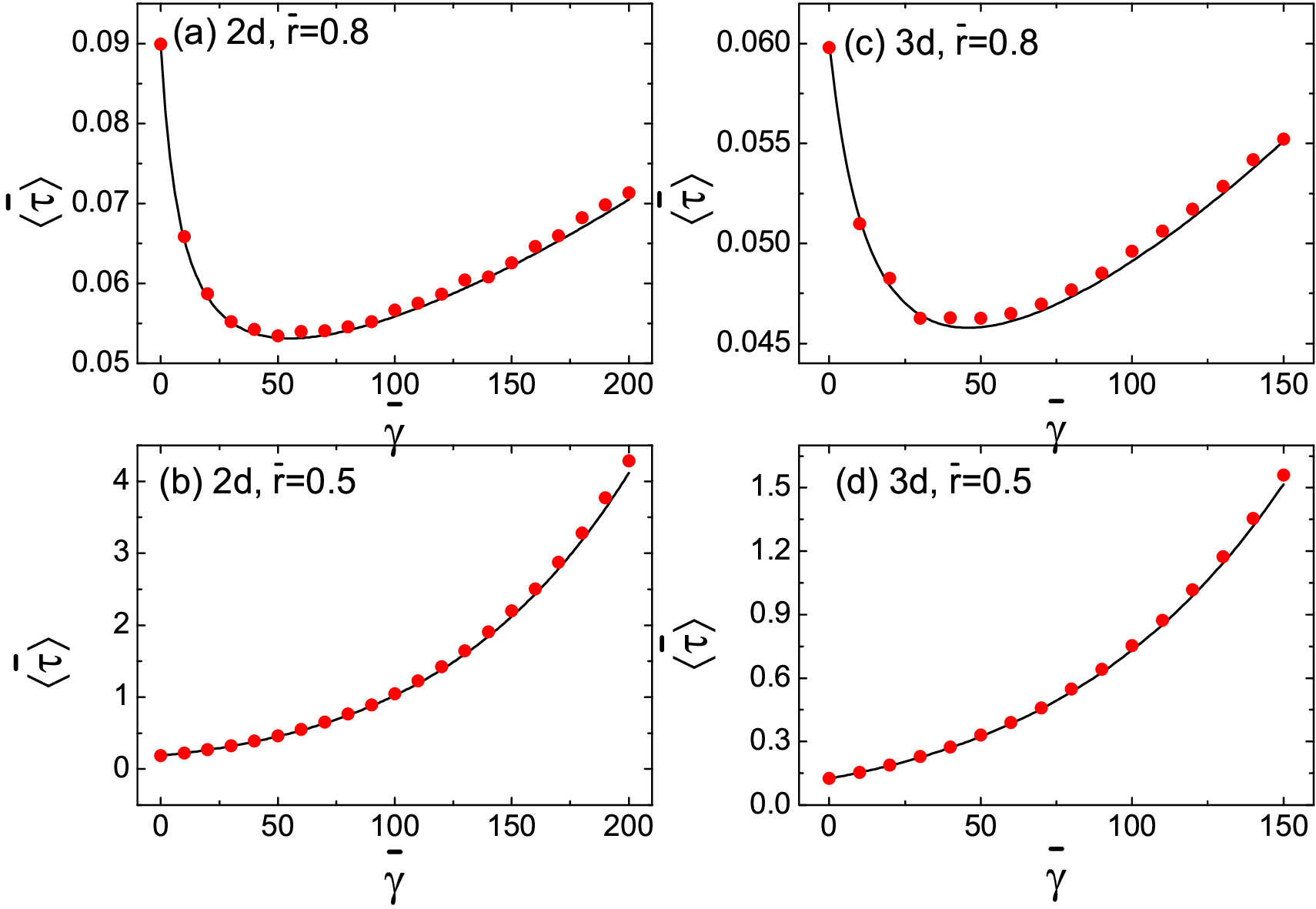}}
	\caption{The dimensionless mean time to adsorption $\langle {\bar \tau } \rangle$ as a function of dimensionless resetting rate $\bar \gamma$ inside 2d and 3d spheres for two different $\bar r$. (a) 2d: $\bar r=0.8$; (b) 2d: $\bar r=0.5$; (c) 3d: $\bar r=0.8$; (d) 3d: $\bar r=0.5$. Comparison is made between simulation (symbols) and theory (lines). In the simulation, we have used the parameters: $D=0.1$, $R=1$ and $dt=10^{-5}$.} \label{fig2}
\end{figure}

In Fig.\ref{fig2}, we plot $\langle {\bar \tau } \rangle$ as a function of $\bar \gamma$ for $d=2$ and $d=3$ and for two different values of $\bar{r}$: $\bar{r}=0.8$ and $\bar{r}=0.5$. When $\bar r$ lies in the domain defined in Eq.\ref{eq2.6}, there exists an optimal $\bar \gamma_{{\rm opt}}$ for which the MTA can be minimized, see Fig.\ref{fig2}(a) ($d=2$ and $\bar{r}=0.8$) and Fig.\ref{fig2}(c) ($d=3$ and $\bar{r}=0.8$). Otherwise, 
$\langle {\bar \tau} \rangle$ show a monotonic increase with $\bar \gamma$, i.e., the resetting prolongs the MTA, see Fig.\ref{fig2}(b) ($d=2$ and $\bar{r}=0.5$) and Fig.\ref{fig2}(d) ($d=3$ and $\bar{r}=0.5$). In order to verify the theoretical results, we have also performed the Langevin dynamics simulations to obtain the MTA. These data (see symbols in Fig.\ref{fig2}) are in good agreement with theory. The details of the numerical simulation are given in the Appendix \ref{app1}.

To determine the optimal resetting rate, we take the derivative of $\langle {\bar \tau } \rangle$ with respect to $\bar{\gamma}$, and the derivative is equal to zero at $\bar{\gamma}=\bar \gamma_{{\rm opt}}$. From Eq.(\ref{eq2.71}), we get the following transcendental equation,
\begin{eqnarray}\label{eq2.72}
\frac{{{{\bar r}^{d/2 - 1}}{I_{d/2 - 1}}\left( {\sqrt {\bar \gamma } } \right)}}{{{I_{d/2 - 1}}\left( {\bar r\sqrt {\bar \gamma } } \right)}} - \frac{{{{\bar r}^{d/2 - 1}}\sqrt {\bar \gamma } \left[ {{I_{d/2 - 2}}\left( {\sqrt {\bar \gamma } } \right) + {I_{d/2}}\left( {\sqrt {\bar \gamma } } \right)} \right]}}{{4{I_{d/2 - 1}}\left( {\bar r\sqrt {\bar \gamma } } \right)}}  \nonumber \\ + \frac{{{{\bar r}^{d/2}}\sqrt {\bar \gamma } {I_{d/2 - 1}}\left( {\sqrt {\bar \gamma } } \right)\left[ {{I_{d/2 - 2}}\left( {\bar r\sqrt {\bar \gamma } } \right) + {I_{d/2}}\left( {\bar r\sqrt {\bar \gamma } } \right)} \right]}}{{4{{\left[ {{I_{d/2 - 1}}\left( {\bar r\sqrt {\bar \gamma } } \right)} \right]}^2}}} =1. \nonumber \\
\end{eqnarray}

In Fig.\ref{fig2.1}, we graphically solve for different values of
$\bar{r}$ in $d=2$. In Fig.\ref{fig3}, we plot $\bar \gamma_{{\rm opt}}$ as a function of $\bar r$ for $d=1$, 2, and 3. $\bar \gamma_{{\rm opt}}$ shows a continuous transition from zero to nonzero value as $\bar r$ passes through the critical value $\bar r_c$. We note that the transition is continuous or second order, such that the domain defined in Eq.\ref{eq2.6} is a sufficient and necessary condition for expediting the MTA via resetting \cite{PhysRevResearch.1.032001}.

\begin{figure}
	\centerline{\includegraphics*[width=0.9\columnwidth]{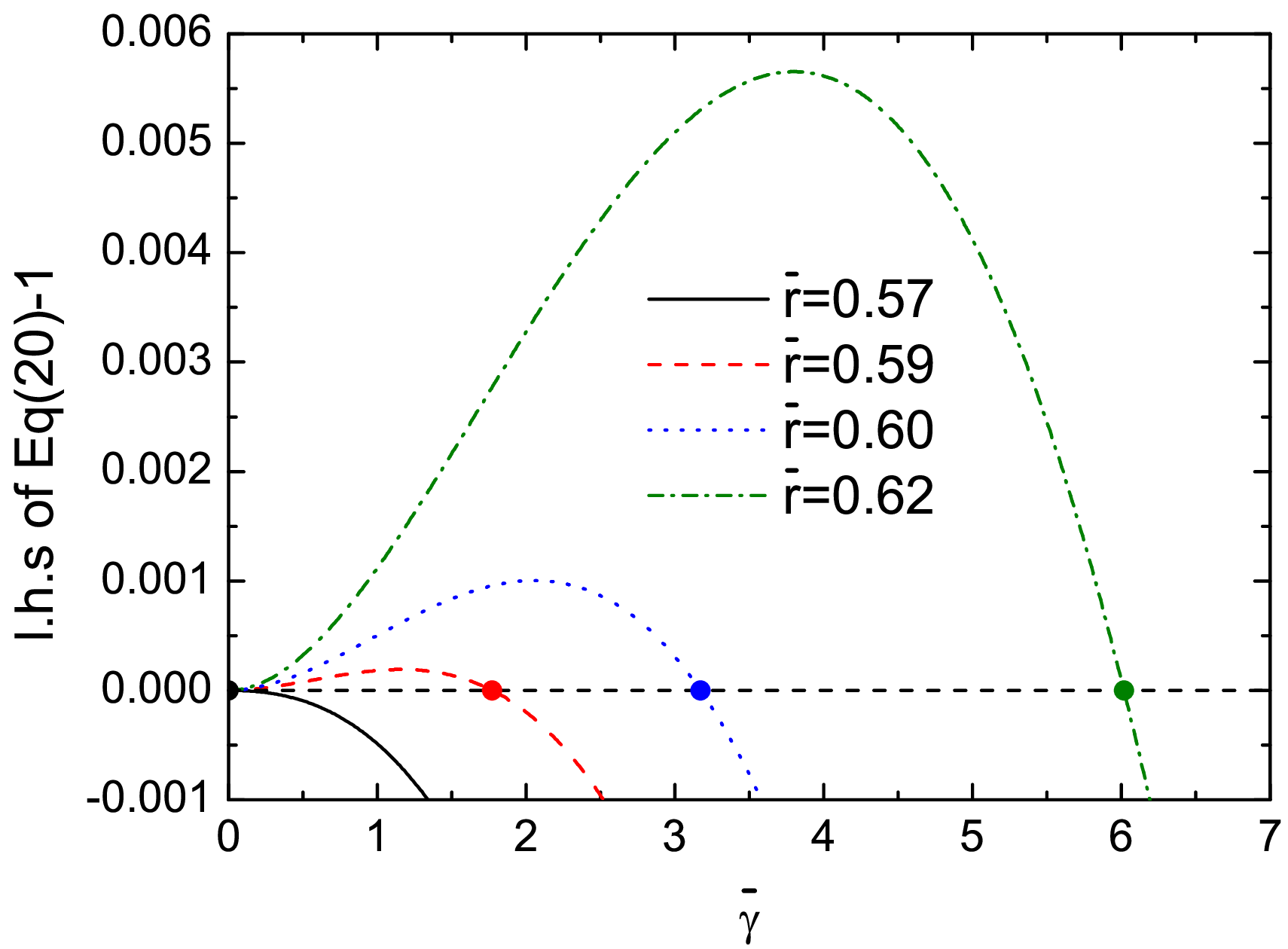}}
	\caption{Graphical solution of Eq.(\ref{eq2.72}) for different values of $\bar{r}$ in $d=2$. The optimal resetting rate $\bar \gamma_{{\rm opt}}$ are highlighted by circles.} \label{fig2.1}
\end{figure}

\begin{figure}
	\centerline{\includegraphics*[width=0.9\columnwidth]{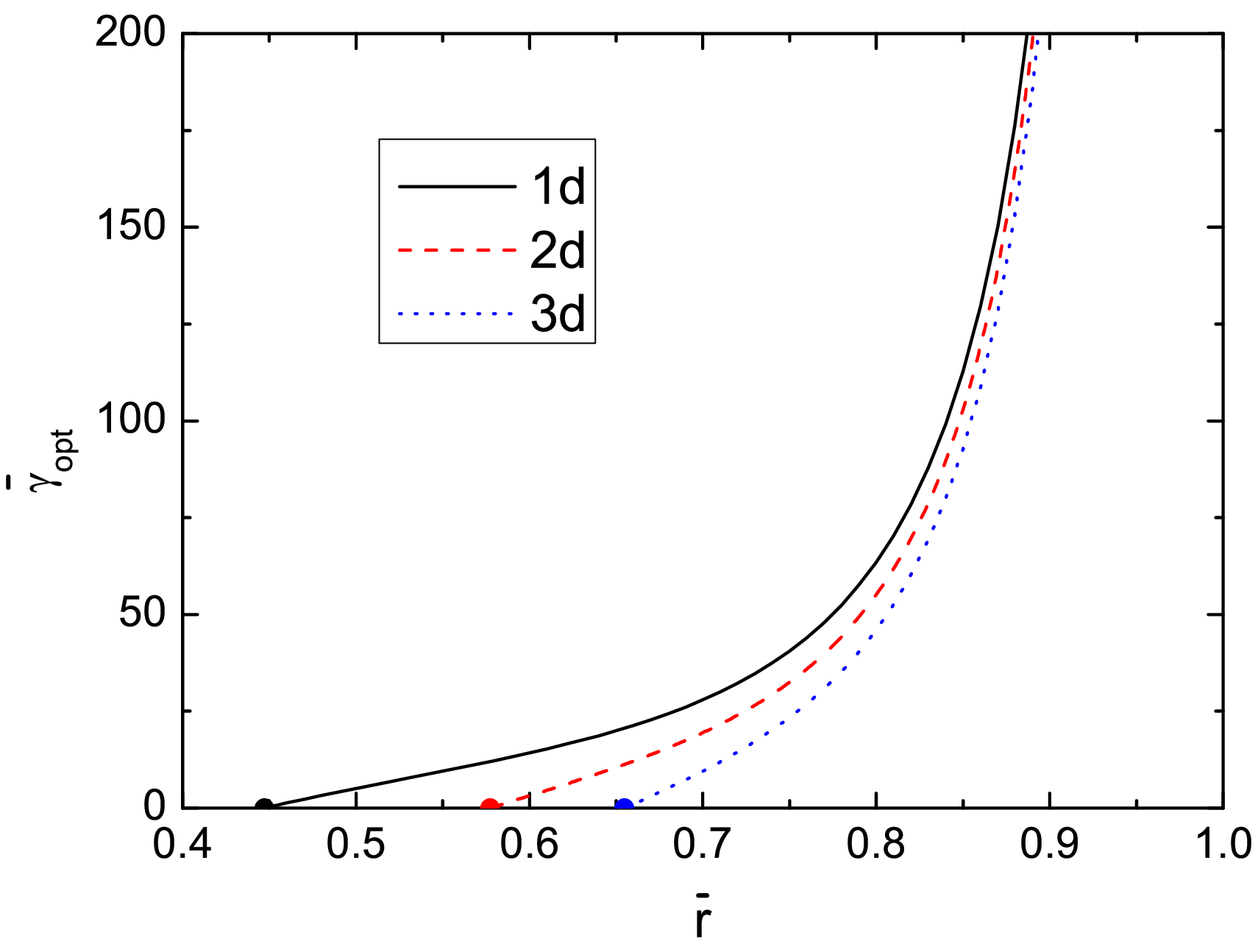}}
	\caption{$\bar \gamma_{{\rm opt}}$ as a function of $\bar r$ for 1d, 2d and 3d spheres.} \label{fig3}
\end{figure}

\section{Results in two $d$-dimensional concentric spheres}
Furthermore, we now turn to a freely diffusive Brownian particle between two $d$-dimensional concentric spheres, starting from a distance $r$ to the center of sphere, $R_1<r<R_2$ with $R_{1(2)}$ being the radius of inner (outer) sphere. The two spherical shells are absorbing boundaries.  See Fig.\ref{figcs} for an illustration. The general solution of $\tilde{Q}_0(s|r)$ is given by Eq.\ref{eq2.2}, imposing on the boundary conditions,
\begin{eqnarray}\label{eq3.10}
\tilde{Q}_0(s|R_1)=\tilde{Q}_0(s|R_2)=0,
\end{eqnarray} 
which yields
\begin{eqnarray}\label{eq3.1}
\tilde Q_0( {s|r} ) = \frac{{{G_{{d}}}( {\alpha,{R_1},r} ) + {G_{{d}}}( {\alpha,r,{R_2}} ) + {G_{{d}}}( {\alpha,{R_2},{R_1}} )}}{{s{G_{{d}}}( {\alpha,{R_2},{R_1}} )}}  \nonumber \\
\end{eqnarray}
where
\begin{eqnarray}\label{eq3.2}
{G_d}( {x,{r_1},{r_2}} ) &=&  {( {{r_1}{r_2}} )^{1 - d/2}} {I_{d/2 - 1}}( {x {r_1}} ){K_{d/2 - 1}}( {x {r_2}} ) \nonumber \\ &-&  {( {{r_1}{r_2}} )^{1 - d/2}} {I_{d/2 - 1}}( {x {r_2}} ){K_{d/2 - 1}}( {x {r_1}} ) \nonumber \\
\end{eqnarray}
and again $\alpha=\sqrt{s/D}$.

\begin{figure}
	\centerline{\includegraphics*[width=0.7\columnwidth]{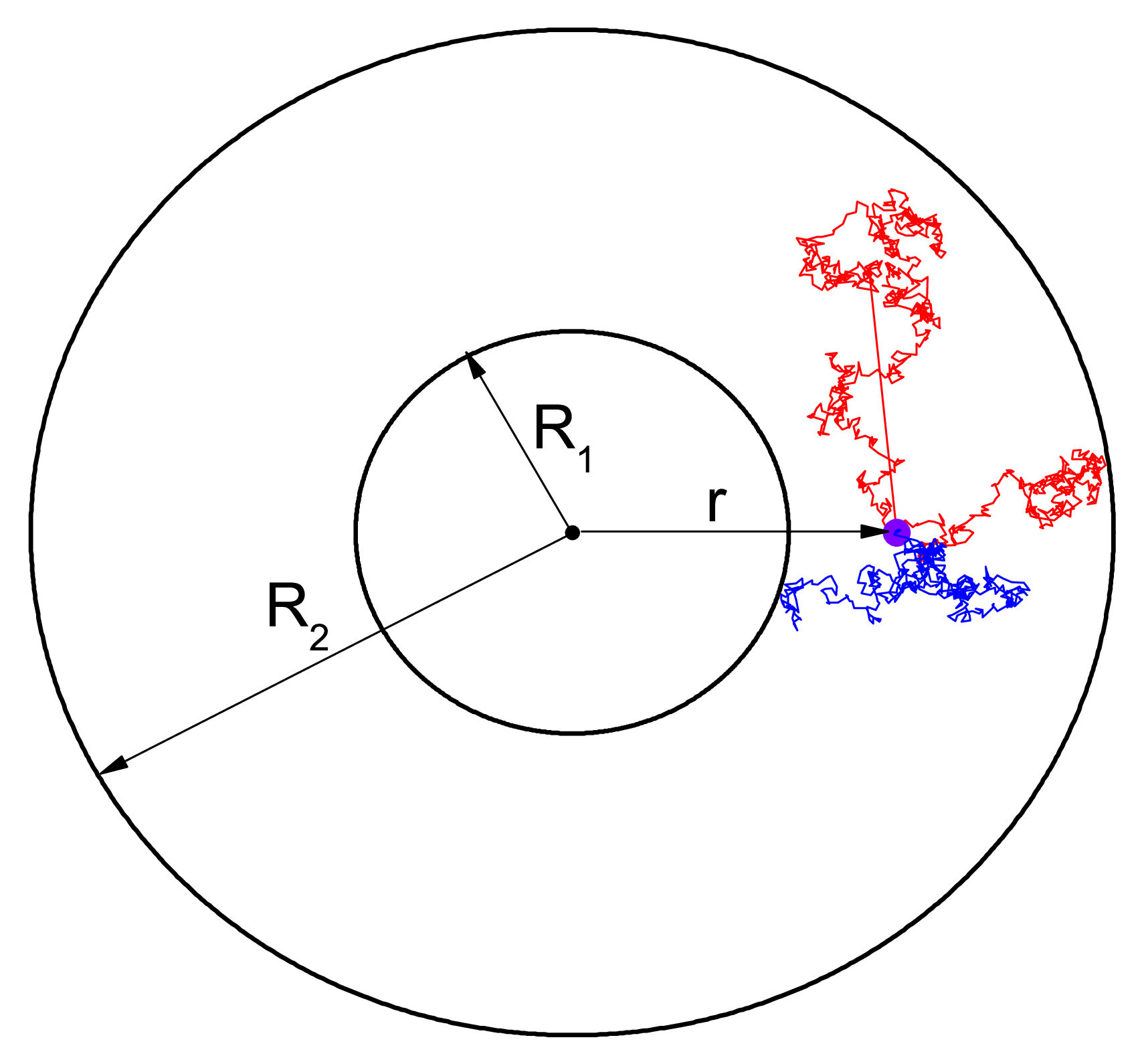}}
	\caption{A Brownian particle diffuses freely between two concentric $d$-dimensional spheres under stochastic resetting. $R_1$ and $R_2$ are the radii of inner sphere and outer sphere, respectively. The particle starts from a distance $r$ ($R_1<r<R_2$) to the centre of sphere. The process is terminated once the particle hits the inner or outer spherical surface.} \label{figcs}
\end{figure}

In terms of Eq.\ref{eq2.01}, a dimensionless MTA in the absence of resetting is given by [see also Chap. 6 of \cite{redner2001guide}]
	\begin{eqnarray}\label{eq3.31}
	\langle {{\bar \tau _0}\left(\bar r \right)} \rangle  =\left\{ \begin{array}{lr}
	\frac{1}{{2d}}\left[ {\frac{{{{\bar r}^{2 - d}} - {c^{2 - d}}}}{{{c^{2 - d}} - 1}}\left( {{c^2} - 1} \right) - \left( {{{\bar r}^2} - c^2} \right)} \right], &d \ne 2\\
	\frac{1}{{2d}}\left[ {\frac{{\ln \bar r}}{{\ln c}}\left( {{c^2} - 1} \right) - \left( {{{\bar r}^2} - 1} \right)} \right],&d = 2
	\end{array} \right. \nonumber \\  
	\end{eqnarray}
where $c=R_1/R_2$ and $\bar r=r/R_2$. Substituting Eq.\ref{eq3.1} into Eq.\ref{eq1.4}, we obtain a dimensionless MTA in the presence of resetting,
\begin{eqnarray}\label{eq3.6}
\langle {\bar \tau} \rangle  =  - \frac{1}{{\bar \gamma }} - \frac{1}{{\bar \gamma }}\frac{{{G_d}\left( {\sqrt {\bar \gamma } ,1,c} \right)}}{{{G_d}\left( {\sqrt {\bar \gamma } ,c,\bar r} \right) + {G_d}\left( {\sqrt {\bar \gamma } ,\bar r,1} \right)}}  . 
\end{eqnarray}

Inspired by the results in spheres, one can speculate that when the particle starts closer to the two absorbing spherical surfaces, the MTA can be optimized via resetting. Otherwise, the resetting is not be beneficial for lowering the MTA. Generally, the acceleration domain is consisted of 
\begin{eqnarray}
\bar r \in (c, \bar r_{c_1}) \cup (\bar r_{c_2},1) .
\end{eqnarray}
However, the derivation of $\bar r_{c_{1,2}}$ is rather complex for general dimensions. Therefore, here we only focus on $d=1$, $d=2$, and $d=3$. This is particularly simple for $d=1$, since the space consists of two disconnected one-dimensional intervals. In such a case, $\bar r_{c_{1,2}}$ can given by the results in Eq.(\ref{eq2.6}) for $d=1$, $\bar r_{c_{1,2}}= \frac{{5(1+c) \mp \sqrt 5 (1-c)}}{{10}}$. Meanwhile, $\bar \gamma{_{\rm opt}}$ shows a continuous transition at $\bar r=\bar r_{c_{1,2}}$, see Fig.\ref{fig4} for the result at $c=0.2$.   

\begin{figure}
	\centerline{\includegraphics*[width=0.9\columnwidth]{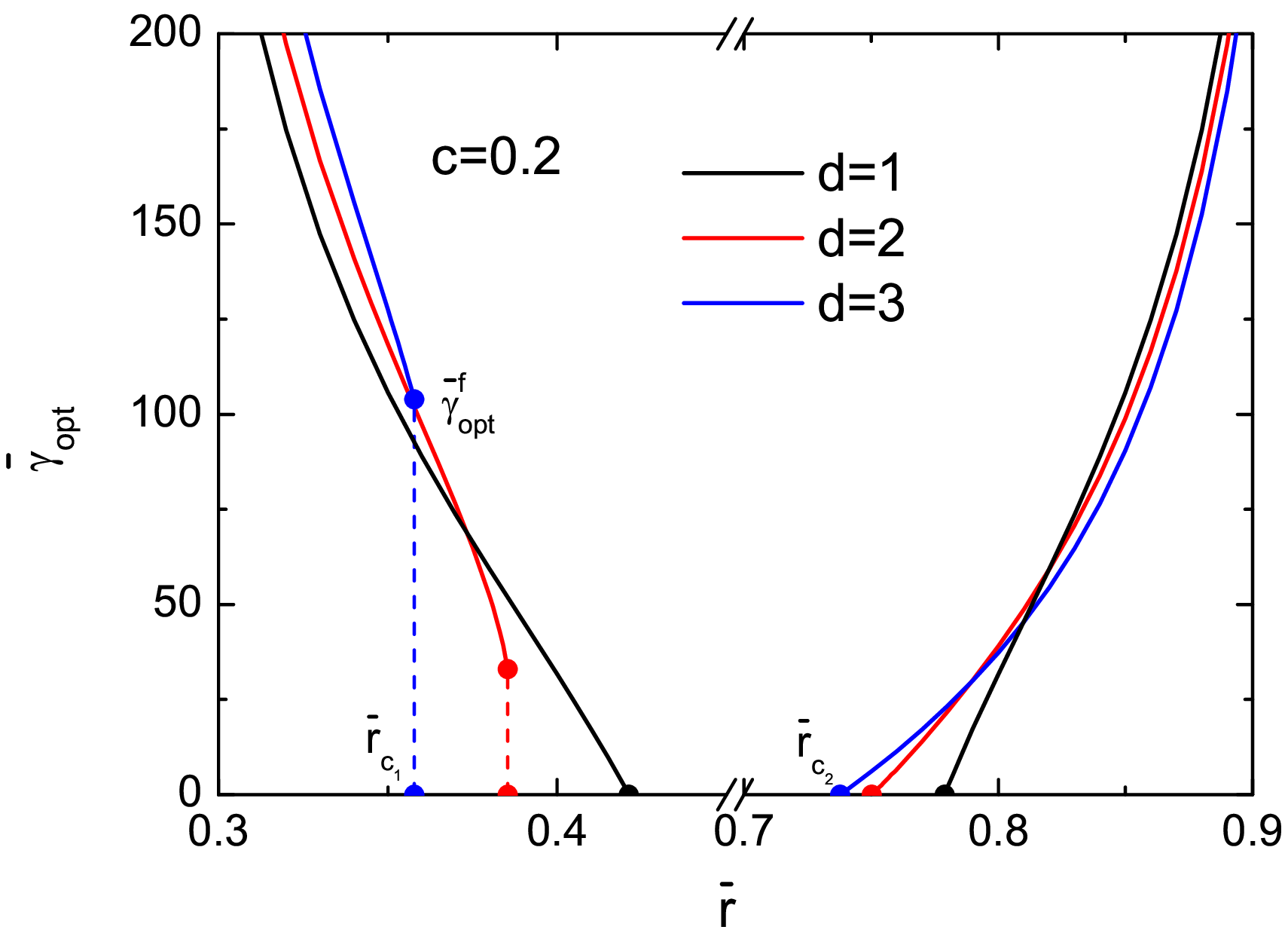}}
	\caption{$\bar \gamma_{{\rm opt}}$ as a function of $\bar r$ for 1d, 2d and 3d concentric spheres. Here $c=0.2$.} \label{fig4}
\end{figure}

Interestingly, for $d=2$ and $d=3$, $\bar \gamma{_{\rm opt}}$ can show a discontinuous transition at $\bar r=\bar r_{c_{1}}$ and a continuous transition at $\bar r=\bar r_{c_{2}}$, see also Fig.\ref{fig4} for the results at $c=0.2$. One can observe that
$\bar \gamma{_{\rm opt}}$ jumps from a finite value $\bar \gamma{_{\rm opt}^f}$ to zero at $\bar r=\bar r_{c_{1}}$. Close to $\bar r_{c_{1}}$, $\bar{\tau}$ shows an inverted $S$-shaped curve (see Fig.\ref{fig5}). This is very analogous to the free energy curve in a first-order phase transition \cite{PhysRevResearch.1.032001}. $\bar{\tau}$ has a local minimum $\bar{\tau}_m$ as $\bar{\gamma}$ varies. Just below $\bar r_{c_{1}}$, $\bar{\tau}_m$ is less than $\bar{\tau}_0$, i.e. the MTA in the absence of resetting, such that a nonzero optimal $\bar{\gamma}$ exists. At $\bar r_{c_{1}}$, $\bar{\tau}_m$ is exactly equal to $\bar{\tau}_0$. Just above $\bar r_{c_{1}}$, $\bar{\tau}_m$ is larger than $\bar{\tau}_0$, and thus a nonzero optimal $\bar{\gamma}$ does not exist.

\begin{figure}
	\centerline{\includegraphics*[width=1.0\columnwidth]{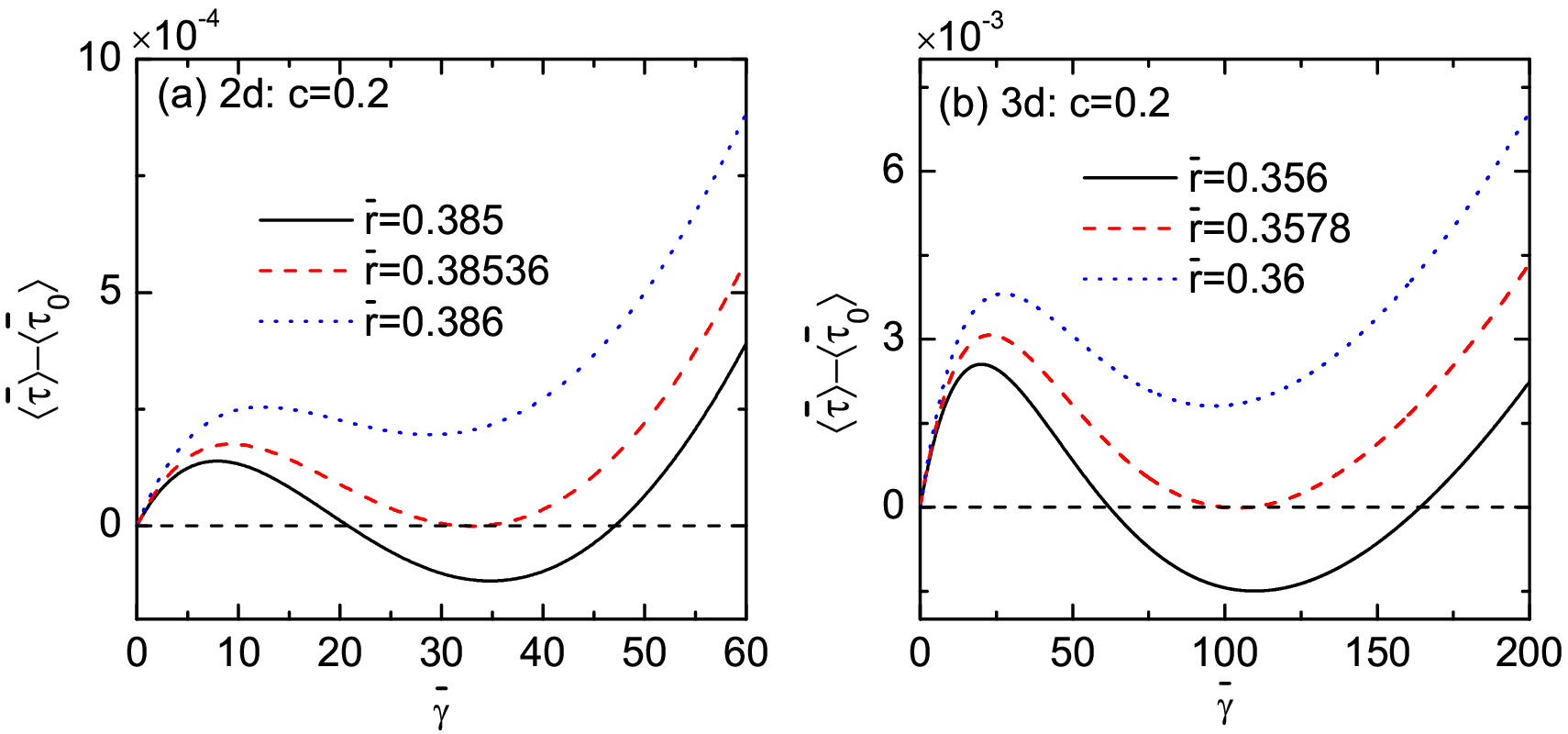}}
	\caption{Shifted dimensionless MTA $\langle \bar \tau \rangle-\langle \bar \tau_0 \rangle$ as a function of $\bar \gamma$ for three different $\bar r$ in 2d (a) and 3d (b) concentric spheres. Here $c=0.2$.} \label{fig5}
\end{figure}

As $c$ increases, the discontinuous phase transition at $\bar r=\bar r_{c_{1}}$ terminates at a tricritical point $(c^*,\bar r^*_{c_{1}})$ and then becomes continuous. In Fig.\ref{fig6}(a) and Fig.\ref{fig6}(b), we show the phase diagram for $d=2$ and $d=3$, respectively. The acceleration domain is located outside of two  transition lines, $\bar r_{c_{1}}(c)$ and $\bar r_{c_{2}}(c)$. The line $\bar r_{c_{2}}(c)$ is of second order, and the line $\bar r_{c_{1}}(c)$ consists of a  first order segment and a second order segment, both ends converge at the tricritical point $(c^*,\bar r^*_{c_{1}})=(0.45,0.58978)$ for $d=2$, and $(c^*,\bar r^*_{c_{1}})=(0.65,0.7383)$ for $d=3$.

\begin{figure}
	\centerline{\includegraphics*[width=1.0\columnwidth]{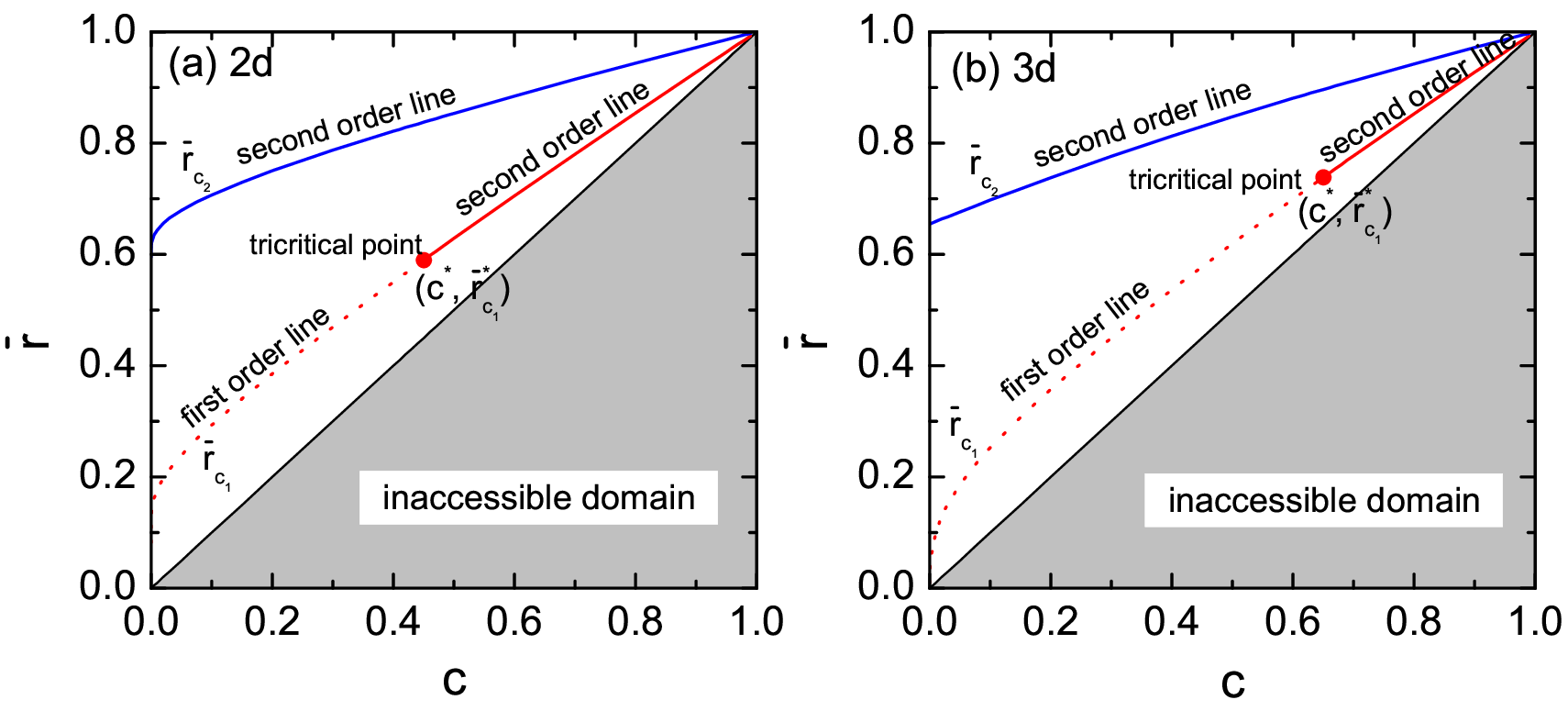}}
	\caption{Phase diagram in the $(c,\bar r)$ plane for 2d (a) and 3d (b) concentric spheres. The first- (in dashed) and second- (in solid) order lines merge at the tricritical points marked by the red circles.} \label{fig6}
\end{figure}

Furthermore, for the second-order transitions, $\bar r_{c_{1,2}}$ can be obtained from the Eq.\ref{eq1.5} \cite{PhysRevResearch.1.032001}. That is to say, $\bar r_{c_{1}}$ can be determined by Eq.\ref{eq1.5} only when $c>c^*$, and $\bar r_{c_{2}}$ can always be determined by Eq.\ref{eq1.5}. While for $c<c^*$, $\bar r_{c_{1}}$ can be obtained by examining Eq.\ref{eq3.6}. After some cumbersome algebra, we find that for $d=2$, $\bar r_{c_{1,2}}$ for are determined by the equation  
\begin{eqnarray}\label{eq3.4}
&& \left( {1 - 4{{\bar r}^2} + 3{{\bar r}^4}} \right){\ln ^2}c + 4{\left( {1 - {c^2}} \right)^2}\ln \bar r\left( {1 + \ln \bar r} \right) \nonumber \\  &+& \left( {1 - {c^2}} \right)\ln c\left[ {4\left( {{{\bar r}^2} - 1} \right) + \left( {3{c^2} + 4{{\bar r}^2} - 5} \right)\ln \bar r} \right]=0.  \nonumber \\  
\end{eqnarray}
In the limit of $c \to 0$, we have  $\bar{r}_{c_2}=\sqrt {1/3}  + \kappa_{2d} {\left| {\ln c} \right|^{ - 1}}$, with $\kappa_{2d}  = \frac{1}{{24}}\left( {16\sqrt 3  - 11\sqrt 3 \ln 3} \right) \approx 0.283$. 

For $d=3$, $\bar r_{c_{1,2}}$ are determined by the equation  
\begin{eqnarray}\label{eq3.5}
&&\left( {7{{\bar r}^4} + 7{{\bar r}^3} - 3{{\bar r}^2} - 3\bar r} \right) + c\left( {10 - 23\bar r - 3{{\bar r}^2} + 7{{\bar r}^3}} \right) \nonumber \\ &+&{c^2}\left( {20 - 23\bar r - 3{{\bar r}^2}} \right) + {c^3}\left( {10 - 3\bar r} \right)   = 0.
\end{eqnarray}
For $c=0$, the solution of Eq.\ref{eq3.5} is $\bar{r}_{c_1}=0$ and $\bar{r}_{c_2}=\sqrt{3/7}$, in accordance with the results of three-dimensional sphere. In the limit of $c \to 0$, we have $\bar{r}_{c_2}=\sqrt {3/7}  + \kappa_{3d} c$, with ${\kappa _{3d}} = \frac{{201\sqrt {21}  - 847}}{{168}} \approx 0.441$.

Finally, we investigate asymptotical behaviors of $\bar{r}_{c_{1}}$ and $\bar{\gamma}_{\rm opt}^f$ in the limit of $c \to 0$, as shown in Fig.\ref{fig7}(a-b) and Fig.\ref{fig7}(c-d) for $d=2$ and $d=3$, respectively. For $d=2$, $\bar{r}_{c_{1}}$ and $\bar{\gamma}_{\rm opt}^f$ show a rather slow change with $c$ as $c \to 0$, which can be fitted as $\bar{r}_{c_{1}}=0.396|\ln c|^{-0.541}$ and $\bar{\gamma}_{\rm opt}^f=20.06 |\ln c|+7.81$. While for $d=3$, $\bar{r}_{c_{1}}$ and $\bar{\gamma}_{\rm opt}^f$ vary with $c$ in a power-law way: $\bar{r}_{c_{1}}=0.532c^{0.352}$ and $\bar{\gamma}_{\rm opt}^f=29.59c^{-0.625}$.

\begin{figure}
	\centerline{\includegraphics*[width=1.0\columnwidth]{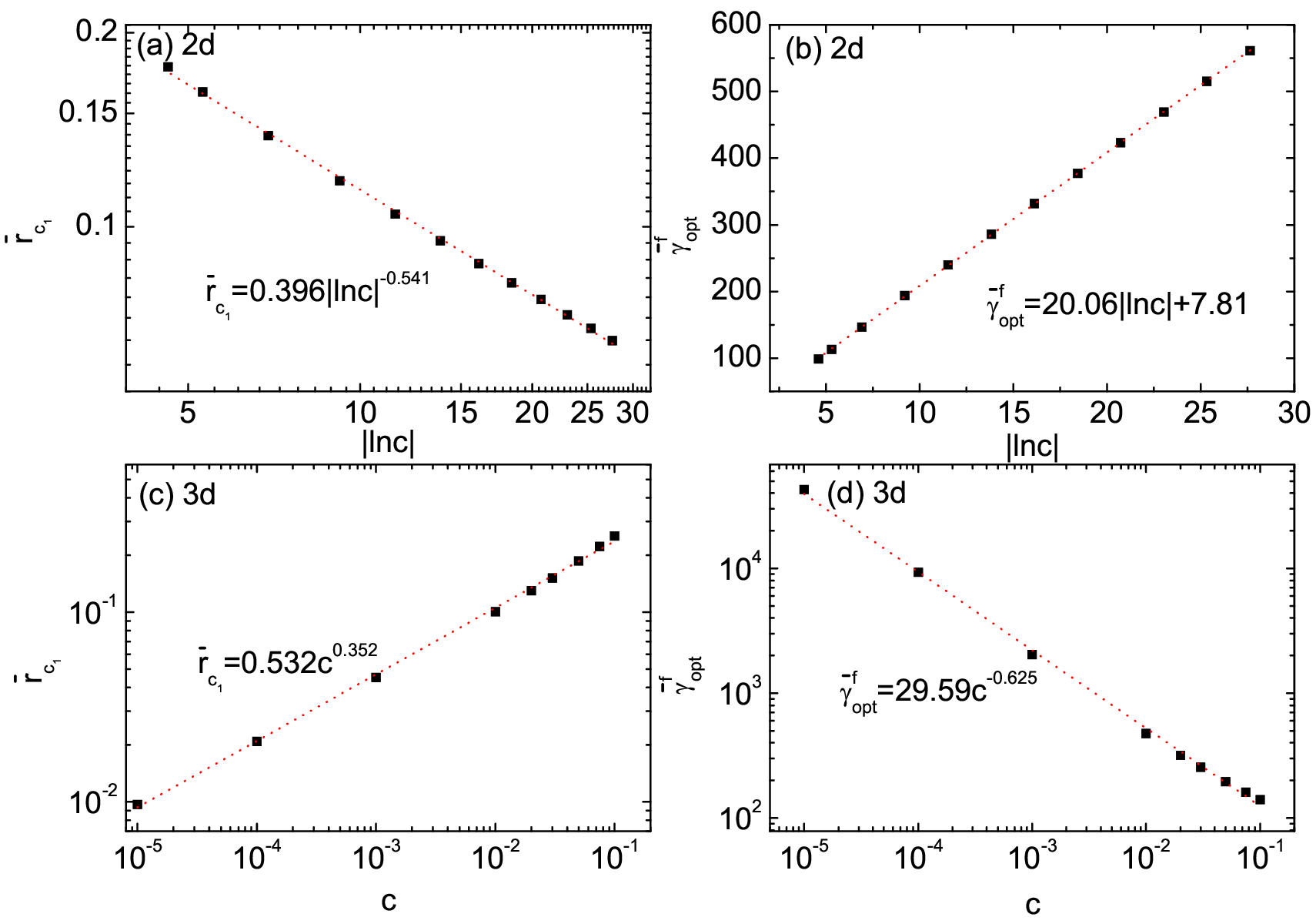}}
	\caption{Asymptotical behaviors of $\bar{r}_{c_{1}}$ and $\bar{\gamma}_{\rm opt}^f$ in the limit of $c \to 0$ for 2d (a-b) and 3d (c-d) concentric spheres. The lines indicate the results of fitting.} \label{fig7}
\end{figure}

\section{Conclusions}

In conclusion, we have investigated the effect of stochastic resetting on first passage properties of a freely diffusing particle inside a sphere or between two concentric spheres, where the spherical surfaces are absorbing boundaries. In the systems, we have observed the so-called ``resetting transition'' when the initial distance $r$ of the particle to the origin is varied. For a $d$-dimensional sphere of radius $R$, there exists a critical value of $r=r_c$, above which the resetting can make the MTA minimal at a nonzero $\gamma$. For $r<r_c$ the MTA increases monotonically  with $\gamma$, and thus the resetting is detrimental to optimization of the MTA. We have derived a general result of $r_c$, see Eq.\ref{eq2.6}, which is a dimension-dependent quantity. For the particle between two $d$-dimensional spheres, there are two critical values of $r=r_{c_{1,2}}$ ($r_{c_1}<r_{c_2}$), for which the MTA can be optimized when $r<r_{c_1}$ or when $r>r_{c_2}$. When the ratio $c$ of the radius of inner sphere to that of outer sphere is less than a critical value, the resetting transition at $r=r_{c_1}$ is of first order, and the  transition at $r=r_{c_2}$ is of second order. Otherwise, the two transitions are of second order.  When the transitions are second order, we have derived the equations for determining these transition points in two dimensions and three dimensions (see Eq.(\ref{eq3.4}) and Eq.(\ref{eq3.5}), respectively).
Furthermore, we have presented the asymptotic analysis for $r_{c_{1,2}}$ and the optimal resetting rate when the first order transition occurs at $r=r_{c_{1}}$ in the limit $c \to 0 $, which shows that the asymptotic behaviors between two dimensions and three dimensions are essentially different.

It is well-known that the first passage of a Brownian particle in the absence of resetting depends highly on the spatial dimensions \cite{redner2001guide,klafter2011first}. From that aspect, our results again unveil the vital role of spatial dimensions on the resetting transition in a spherically symmetric bounded domain. In the future, it is worth considering cases where the particle experiences a radial potential drift or in the presence of active driving. It would also be interesting to study the resetting Brownian motion in a bounded domain with a more complex boundary, such as a small absorbing window on the otherwise reflecting boundary. These extensions may provide a valuable understanding for controlling intracellular transport such as proteins and other molecular products moving to their correct location in a plasma membrane, and diffusion-limited reactions \cite{redner2001guide,RevModPhys.85.135}.

\appendix
\section{Simulation details}\label{app1}
In the resetting Brownian motion model, the position $\bme{x}(t)$ of a Brownian particle is reset to the origin $\bme{x}(0)$ randomly in time according to a Poission process with a constant rate $\gamma$. In a time interval $dt$, the position follows the stochastic Langevin dynamics
\begin{eqnarray}\label{eqa1}
  {x_i}\left( {t + dt} \right) = \left\{ \begin{array}{llc}
  {x_i}\left( t \right) + \sqrt {2 D dt} {\xi _i}(t), & \,  {\rm{with}} \, {\rm{prob.}} \, & 1-\gamma dt,  \\
  {x_i}\left( 0 \right), &  \,  {\rm{with}} \, {\rm{prob.}}  \, & \gamma dt,  \\ 
  \end{array}  \right. \nonumber \\
\end{eqnarray}
where $x_i$ is the $i$th component of $\bme{x}$ in the Cartesian coordinate system, $D$ is the diffusion coefficient, and ${\xi _i}(t)$ is a Gaussian white noise with mean zero and variance given by $\left\langle {{\xi _i}\left( t \right){\xi _j}\left( {t'} \right)} \right\rangle  = {\delta _{ij}}\delta \left( {t - t'} \right)$. The noise is generated via the Box-Muller algorithm \cite{Box_Muller}.

In the beginning of simulation, the Brownian particle is placed inside a bounded domain, $\bme{x}(0) \in \Omega$. The position of the particle is updated in terms of Eq.(\ref{eqa1}), and the simulation ends when the particle crosses the boundary $\partial \Omega$ of the domain. The time to absorption at the boundary is given by
\begin{eqnarray}\label{eqa2}
{\tau _0} = \inf \left\{ {t:\bme{x}(t) \notin \Omega } \right\}.
\end{eqnarray}
In the simulation, we have set $D=0.1$ and $dt=10^{-5}$. For each datum, we have performed $10^5$ independent simulations to obtain the mean time to adsorption $\langle \tau _0 \rangle$.

\begin{acknowledgments}
This work is supported by the National Natural Science Foundation of China (Grants No. 11875069 and No 61973001) and the Key Scientific Research Fund of Anhui Provincial Education Department under (Grant No. KJ2019A0781).
\end{acknowledgments}


\end{document}